\documentclass{article}
\usepackage{amssymb}

\input{tcilatex}
\begin{document}

\author{S. I. Santos J\'{u}nior \\
Department of Physics\\
Centre for Technological Sciences-UDESC\\
Joinville 89223-100, Santa Catarina, Brazil.\\
e-mail: samuel.isidoro.santos@gmail.com \and J. G. Cardoso \\
Department of Mathematics\\
Centre for Technological Sciences-UDESC\\
e-mail: jorge.cardoso@udesc.br}
\title{Wave Equations for Classical Two-Component Proca Fields in Curved
Spacetimes with Torsionless Affinities}
\date{ }
\maketitle

\begin{abstract}
The world formulation of the full theory of classical Proca fields in
generally relativistic spacetimes is reviewed. Subsequently the entire set
of field equations is transcribed in a straightforward way into the
framework of one of the Infeld-van der Waerden formalisms. Some well-known
calculational techniques are then utilized for deriving the wave equations
that control the propagation of the fields allowed for. It appears that no
interaction couplings between such fields and electromagnetic curvatures are
ultimately carried by the wave equations at issue. What results is, in
effect, that the only interactions which occur in the theoretical context
under consideration involve strictly Proca fields and wave functions for
gravitons.
\end{abstract}

\section{Introduction}

Traditionally, the Infeld-van der Waerden $\gamma \varepsilon $-formalisms
[1] constitute the classical two-component spinor framework for general
relativity. The construction of these formalisms was primarily aimed at
exhibiting an elementary description of the dynamics of Dirac fields in
generally relativistic spacetimes. Such formalisms had been designed
originally much earlier than the achievement of the definitive conditions
for a curved space to admit spinor structures locally [2-4]. The legitimacy
of the procedures for building up them relies crucially upon the existence
of sets of Hermitian connecting objects at non-singular spacetime points.
Their affine prescriptions were formally shaped upon the ones that occur in
the realm of general relativity. Thus, the generalized Weyl gauge group [5]
is taken to operate on spin spaces set up locally in a way that does not
depend at all upon the action of manifold mapping groups. Loosely speaking,
all curvature spinors arise from the decomposition of mixed world-spin
quantities that result from the action of torsionless covariant derivative
commutators on arbitrary spin vectors [1, 6]. The $\gamma \varepsilon $%
-formalisms were extensively utilized over the years by many authors in
several different ways [6-29], noticeably enough, to reconstruct some
classical generally relativistic structures and to transcribe classification
schemes for world curvature tensors. Notwithstanding the fact that the
construction of curvature spinors is implicitly carried by the formalisms,
the spin curvatures that occur in the classification schemes and some of the
spinor structures mentioned above were obtained in an artificial way by
carrying out straightforward spinor translations of Riemann and Weyl
tensors. A fairly complete algebraic description of the affine and curvature
structures tied in with the formalisms is supplied in Refs. [30-32].

The most striking physical feature of the $\gamma \varepsilon $-framework
lies over the result that any curvature spinors are expressed as sums of
purely gravitational and electromagnetic contributions which produce, in an
inextricably geometric way, the occurrence of wave functions for gravitons
and photons of both handednesses. The gravitational contributions for the $%
\varepsilon $-formalism were utilized in Refs. [18, 19] to support a spinor
translation of Einstein's equations. It had been established a little
earlier [20] that any of them should show up as a spinor pair which must be
associated to the irreducible decomposition of a Riemann tensor. Any
gravitational wave functions for either formalism are defined as totally
symmetric curvature pieces that occur in spinor decompositions of Weyl
tensors [18]. On the other hand, each electromagnetic curvature contribution
emerges as a pair of suitably contracted pieces\ which enter the spinor
representation of a locally defined Maxwell bivector [30]. The work of Ref.
[4] gives a rough description of the propagation of gravitons for the $%
\varepsilon $-formalism together with a derivation of the patterns for their
interactions with external electromagnetic fields. In Refs. [30, 32], the
full $\gamma \varepsilon $-description of the propagation of spin curvatures
in vacuum is brought out. It thus appears that the couplings between
gravitons and photons are strictly borne in both formalisms by the wave
equations that govern the electromagnetic propagation. The propagation of
gravitons in the presence of arbitrary sources is described in Ref. [33]
where a somewhat important condition on the first covariant derivative of
energy-momentum tensors is deduced. A specialization of this description for
the particular case of sources coming from eletromagnetic curvatures, is
given in Ref. [26]. The work of Ref. [27] touches upon an interesting
situation concerning the occurrence of geometric sources in the field
equations for Infeld-van der Waerden photons. In Ref. [30], it was suggested
for the first time that a description of some of the physical properties of
the cosmic microwave background could be achieved by looking at the
propagation in Friedmann-like conformally flat spacetimes of electromagnetic
curvatures. A notable class of conformally flat spacetimes which admit
decomposable Christoffel connexions, was considered in Ref. [21] in
conjunction with a derivation of the corresponding spin-affine and curvature
configurations for the $\gamma $-formalism. It was shown thereabout that the
whole derivation can actually be implemented only if a specific constancy
property is imposed on one of the spin densities borne by the expression for
a characteristic $\gamma $-metric function. Explicit expressions for the
gravitational spinors of those spacetimes were then derived. A detailed
description of the interaction couplings that take place in the formulation
of Dirac's theory in curved spacetimes has likewise been given [22]. This
latter work has really made up the original description of Dirac fields as
given by Infeld and van der Waerden.

In the present work, we exhibit the formulation of the theory of classical
Proca fields within the framework of the $\gamma $-formalism. The theory of
spin densities and the gauge transformations inherently borne by the $%
\varepsilon $-formalism [30, 31] will not be exhibited by this point since
the $\varepsilon $-counterparts of our key developments do not bring forth
any further formal insight. The spinor field equations are obtained out of
transcribing directly the statements that make out the world version of the
theory. Some well-known calculational techniques are then utilized for
deriving the wave equations that control the propagation of the fields taken
into consideration. Indeed, these techniques are just the same as the ones
employed in Refs. [23, 30] for obtaining the typical wave equations of the
entire $\gamma \varepsilon $-framework. Hence, no interaction couplings
between Proca fields and electromagnetic curvatures are ultimately carried
by the resulting wave equations. What comes about is, in effect, that the
only interactions which occur in the theoretical context being considered
involve strictly Proca fields and wave functions for gravitons. One of our
motivations for elaborating upon the situation entertained herein is related
to the absence from the literature of any systematic two-component
description of the propagation of massive spinning bosons in generally
relativistic spacetimes. In our view, it might be worthwhile to work out
such a massive case towards exhibiting the patterns of the couplings that
should arise in the pertinent context. It is from this fact that the main
physical aim of our paper stems.

We will adopt the notation adhered to in Ref. [30] except that spacetime
components will now be labelled by lower-case Greek letters. Kernel letters
for world and spin quantities will broadly appear as Greek and Latin
letters. In particular, we denote as $x^{\mu }$ some local coordinates on a
spacetime $\mathfrak{M}$ equipped with a torsionless covariant derivative
operator $\nabla _{\mu }$. A world metric tensor $g_{\mu \nu }$ on $%
\mathfrak{M}$ presumably bears the local signature $(+---)$. We thus require 
$g_{\mu \nu }$ to fulfill at the outset the metric compatibility condition
of general relativity%
\begin{equation*}
\nabla _{\mu }g_{\lambda \sigma }=0,
\end{equation*}%
which means that we shall allow for the (unique) Levi-Civita connection
associated to $\nabla _{\mu }$. The partial derivative operator for $x^{\mu
} $ is denoted by $\partial _{\mu }$, and the Riemann tensor of $\nabla
_{\mu } $ is written as $R_{\mu \nu \lambda \sigma }$. Our sign convention
for the respective Ricci tensor $R_{\mu \nu }$ is the same as the one
adopted in Ref. [4]. The determinant of $g_{\mu \nu }$ and the covariant
alternating world density in $\mathfrak{M}$ will especially be denoted as $%
\mathfrak{g}$ and $\mathfrak{\epsilon }_{\mu \nu \lambda \sigma }$,
respectively. We shall use the primed-unprimed index notation of Ref. [4]
upon dealing with conjugate spinor components. World indices all range over
the four values $0,1,2,3$ whereas spinor indices take either the values $0,1$
or $0^{\prime },1^{\prime }$. We will utilize the convention according to
which the effect on any index block of the actions of the symmetry and
antisymmetry operators is indicated by surrounding the indices singled out
with round and square brackets, respectively. A horizontal bar lying over a
kernel letter will sometimes be used to denote the operation of complex
conjugation. Further conventions will be explained in due course.

Our outline has been set as follows. For the sake of consistency, the world
version of the Proca theory is formulated in Section 2 on the basis of the
standard least-action principle for classical fields in curved spacetimes
[34]. Section 3 brings out the overall system of spinor field equations. In
Section 4, we carry out the derivation of our wave equations. Some remarks
on our work are made in Section 5. The calculational techniques referred to
above shall be taken for granted from the beginning.

\section{World theory}

The least-action principle for the Proca theory in $\mathfrak{M}$ is written
as%
\begin{equation}
\delta S=\delta \int_{\Omega }\mathcal{L}\sqrt{-\mathfrak{g}}d^{4}x=0,
\label{e1}
\end{equation}%
where $\mathcal{L}$ denotes the Lagrangian density%
\begin{equation}
\mathcal{L}=-\frac{1}{4}f^{\mu \nu }f_{\mu \nu }+m^{2}A^{\mu }A_{\mu },
\label{e2}
\end{equation}%
which carries the Proca bivector%
\begin{equation}
f_{\mu \nu }=2\partial _{\lbrack \mu }A_{\nu ]}=2\nabla _{\lbrack \mu
}A_{\nu ]},  \label{e3}
\end{equation}%
with $A_{\mu }$ and $m$ being a Proca potential and the mass of $f_{\mu \nu
} $. Usually, the variation $\delta $ bears linearity and obeys the Leibniz
rule, in addition to being defined so as to commute with partial derivatives
and integrations. The integral of Eq. (\ref{e1}) is taken over a volume $%
\Omega $ in $\mathfrak{M}$ whose closure is compact, and%
\begin{equation}
d^{4}x=\frac{1}{4!}\mathfrak{\epsilon }_{\mu \nu \lambda \sigma }dx^{\mu
}\wedge dx^{\nu }\wedge dx^{\lambda }\wedge dx^{\sigma }  \label{e4}
\end{equation}%
defines an elementary volume density in $\Omega $, with the symbol
\textquotedblleft $\wedge $\textquotedblright\ thus denoting the wedge
product.

With the help of Eqs. (\ref{e2}) and (\ref{e3}), we can rewrite the
statement (\ref{e1}) as%
\begin{equation}
\delta S=\int_{\Omega }(-f^{\mu \nu }\partial _{\mu }\delta A_{\nu
}+m^{2}A^{\nu }\text{$\delta $}A_{\nu })\sqrt{-\mathfrak{g}}d^{4}x=0,
\label{e5}
\end{equation}%
where $\delta A_{\nu }$ is taken as an arbitrary covariant quantity in $%
\Omega $ that vanishes on the boundary $\partial \Omega $ of $\Omega $.
Hence, performing an integration by parts in (\ref{e5}), yields%
\begin{align}
& \int_{\Omega }[\frac{1}{\sqrt{-\mathfrak{g}}}\partial _{\mu }(\sqrt{-%
\mathfrak{g}}f^{\mu \nu })+m^{2}A^{\nu }]\sqrt{-\mathfrak{g}}\delta A_{\nu
}d^{4}x  \notag \\
& -\int_{\partial \Omega }f^{\mu \nu }\sqrt{-\mathfrak{g}}\delta A_{\nu
}d^{3}x_{\mu }=0,  \label{e6}
\end{align}%
with%
\begin{equation}
d^{3}x_{\mu }=\frac{1}{3!}\mathfrak{\epsilon }_{\mu \nu \lambda \sigma
}dx^{\nu }\wedge dx^{\lambda }\wedge dx^{\sigma },  \label{e7}
\end{equation}%
whence we can write down the field equations%
\begin{equation}
\frac{1}{\sqrt{-\mathfrak{g}}}\partial _{\mu }(\sqrt{-\mathfrak{g}}f^{\mu
\nu })+m^{2}A^{\nu }=0,  \label{e8}
\end{equation}%
which amount to the same thing as%
\begin{equation}
\nabla _{\mu }f^{\mu \nu }+m^{2}A^{\nu }=0.  \label{e9}
\end{equation}

Equations (\ref{e9}) constitute the first world half of Proca's theory in $%
\mathfrak{M}$. The second half comes into play as the Bianchi identity%
\begin{equation}
\nabla _{\lbrack \mu }f_{\lambda \sigma ]}=0,  \label{e10}
\end{equation}%
which can be reexpressed as%
\begin{equation}
\nabla ^{\mu }{}f_{\mu \nu }^{\ast }=0,  \label{e11}
\end{equation}%
where 
\begin{equation}
f_{\mu \nu }^{\ast }=\frac{1}{2}\sqrt{-\mathfrak{g}}\mathfrak{\epsilon }%
_{\mu \nu \lambda \sigma }f^{\lambda \sigma }  \label{e12}
\end{equation}%
is the dual bivector of $f_{\mu \nu }$. It follows that, by taking the
covariant divergence of the left-hand side of Eq. (\ref{e9}), likewise
implementing the commutator expansion%
\begin{equation}
\lbrack \nabla _{\mu },\nabla _{\nu }]f^{\mu \nu }=R_{\mu \nu \lambda
}{}^{\mu }f^{\lambda \nu }+R_{\mu \nu \lambda }{}^{\nu }f^{\mu \lambda
}=2R_{\mu \nu }f^{\mu \nu }\equiv 0,  \label{e14}
\end{equation}%
we promptly arrive at%
\begin{equation}
\nabla _{\mu }A^{\mu }=0.  \label{e15}
\end{equation}%
A similar procedure involves inserting (\ref{e3}) into (\ref{e9}) to get%
\begin{equation}
(\square +m^{2})A_{\mu }-\nabla _{\lambda }\nabla _{\mu }A^{\lambda }=0,
\label{e16}
\end{equation}%
whence, making use of the equality%
\begin{equation}
\nabla _{\lambda }\nabla _{\mu }A^{\lambda }=\nabla _{\mu }(\nabla _{\lambda
}A^{\lambda })+R_{\lambda \mu \sigma }{}^{\lambda }A^{\sigma },  \label{e17}
\end{equation}%
and calling upon (\ref{e15}), we end up with the wave equation%
\begin{equation}
(\square +m^{2})A_{\mu }+R_{\mu }{}^{\lambda }A_{\lambda }=0.  \label{e18}
\end{equation}

\section{Spinor field equations}

By definition, any Proca wave functions enter formal bivector expansions like

\begin{equation}
\sigma _{AA^{\prime }}^{\mu }\sigma _{BB^{\prime }}^{\nu }f_{\mu \nu
}=f_{AA^{\prime }BB^{\prime }}=\gamma _{A^{\prime }B^{\prime }}\psi
_{AB}+\gamma _{AB}\psi _{A^{\prime }B^{\prime }},  \label{e19}
\end{equation}%
with%
\begin{equation}
\psi _{AB}=\frac{1}{2}f_{ABC^{\prime }}{}^{C^{\prime }}=\psi _{(AB)},\text{ }%
\psi _{A^{\prime }B^{\prime }}=\frac{1}{2}f_{A^{\prime }B^{\prime
}C}{}^{C}=\psi _{(A^{\prime }B^{\prime })},  \label{e20}
\end{equation}%
and $(\gamma _{AB},$ $\gamma _{A^{\prime }B^{\prime }})$ being a pair of
covariant metric spinors for the $\gamma $-formalism. The $\sigma $-symbols
carried by Eq. (\ref{e19}) are some appropriate Hermitian connecting
objects, which supposedly fulfill the covariant constancy requirement (see,
for instance, Ref. [32])%
\begin{equation}
\nabla _{\mu }\sigma _{BB^{\prime }}^{\lambda }=0.  \label{e21}
\end{equation}%
Accordingly, either of $\psi _{AB}$ and $\psi _{A^{\prime }B^{\prime }}$ is
a massive spin-one uncharged field that represents locally the six degrees
of freedom of $f_{\mu \nu }$ in $\mathfrak{M}$. The corresponding
field-potential relationships are given by%
\begin{equation}
\psi _{AB}=-\nabla _{(A}^{C^{\prime }}A_{B)C^{\prime }},\text{ }\psi
_{A^{\prime }B^{\prime }}=-\nabla _{(A^{\prime }}^{C}A_{B^{\prime })C}
\label{e22}
\end{equation}%
and%
\begin{equation}
\psi {}^{AB}=\nabla _{C^{\prime }}^{(A}A^{B)C^{\prime }},\text{ }\psi
^{A^{\prime }B^{\prime }}=\nabla _{C}^{(A^{\prime }}A^{B^{\prime })C}.
\label{e23}
\end{equation}

In passing, we point out that, in deriving Eqs. (\ref{e22}) and (\ref{e23}),
it may be necessary to implement the Infeld-van der Waerden eigenvalue
equations [6, 31]%
\begin{equation}
\nabla _{\mu }\gamma _{BC}=i\beta _{\mu }\gamma _{BC},\text{ }\nabla _{\mu
}\gamma ^{BC}=-i\beta _{\mu }\gamma ^{BC},  \label{e24}
\end{equation}%
as well as their complex conjugates. The quantity $\beta _{\mu }$ amounts to
the world vector%
\begin{equation}
\beta _{\mu }=\nabla _{\mu }\Phi +2\Phi _{\mu },  \label{e25}
\end{equation}%
which is invariant under the action of the Weyl gauge group [1, 6], with $%
\Phi $ and $\Phi _{\mu }$ being, respectively, the polar argument of the
independent component of $\gamma _{AB}$ and a $\gamma $-formalism
electromagnetic potential. It is useful to introduce the Maxwell bivector
associated to $\Phi _{\mu }$. We have, in effect,%
\begin{equation}
F_{\mu \nu }=2\partial _{\lbrack \mu }\Phi _{\nu ]}=2\nabla _{\lbrack \mu
}\Phi _{\nu ]}.  \label{e26}
\end{equation}%
The spinor decomposition of $F_{\mu \nu }$ takes up the wave functions $\phi
_{AB}$ and $\phi _{A^{\prime }B^{\prime }}$ which thus supply dynamical
states for Infeld-van der Waerden photons in $\mathfrak{M}$. These wave
functions essentially constitute the electromagnetic curvature of $\mathfrak{%
M}$, thereby being deeply rooted into the geometric structure of $\mathfrak{M%
}$. It is obvious that the geometric field-potential relationships may right
away be attained from (\ref{e22}) and (\ref{e23}) by making trivial
replacements of kernel letters. Such relationships shall be utilized later
in Section 4.

The first spinor half of Proca's theory arises here from the two-component
transcription of Eq. (\ref{e9}) whence, by invoking (\ref{e21}), we obtain
the field equations\footnote{%
The symbol \textquotedblleft c.c.\textquotedblright\ will henceforth denote
an overall complex conjugate piece.}%
\begin{equation}
\nabla ^{AA^{\prime }}(\gamma _{A^{\prime }B^{\prime }}\psi _{AB}+\text{c.c.}%
)+m^{2}A_{BB^{\prime }}=0.  \label{e27}
\end{equation}%
The second half now consists of the statements

\begin{equation}
\nabla ^{AA^{\prime }}f_{AA^{\prime }BB^{\prime }}^{\ast }=i\nabla
^{AA^{\prime }}\left( \gamma _{AB}\psi _{A^{\prime }B^{\prime }}-\text{c.c.}%
\right) =0,  \label{e28}
\end{equation}%
which effectively account for the dual expansion%
\begin{equation}
\sigma _{AA^{\prime }}^{\mu }\sigma _{BB^{\prime }}^{\nu }f_{\mu \nu }^{\ast
}=i(\gamma _{AB}\psi _{A^{\prime }B^{\prime }}-\gamma _{A^{\prime }B^{\prime
}}\psi _{AB}).  \label{e29}
\end{equation}%
Of course, Eq. (\ref{e28}) may be reset as the Hermitian configuration%
\begin{equation}
\nabla ^{AA^{\prime }}\left( \gamma _{A^{\prime }B^{\prime }}\psi
_{AB}\right) =\nabla ^{AA^{\prime }}\left( \gamma _{AB}\psi _{A^{\prime
}B^{\prime }}\right) .  \label{e30}
\end{equation}%
We should also observe that the pattern of Eq. (\ref{e29}) oftenly emerges
from the combination of (\ref{e19}) with the alternating expansion%
\begin{equation}
\sqrt{-\mathfrak{g}}\mathfrak{\epsilon }_{AA^{\prime }BB^{\prime }CC^{\prime
}DD^{\prime }}=i(\gamma _{AC}\gamma _{BD}\gamma _{A^{\prime }D^{\prime
}}\gamma _{B^{\prime }C^{\prime }}-\gamma _{AD}\gamma _{BC}\gamma
_{A^{\prime }C^{\prime }}\gamma _{B^{\prime }D^{\prime }}).  \label{e31}
\end{equation}

Typically, the entire Proca theory in $\mathfrak{M}$ is written out
explicitly as the field equations 
\begin{equation}
\nabla _{B^{\prime }}^{A}\psi _{AB}+\frac{1}{2}m^{2}A_{BB^{\prime }}-i\beta
_{B^{\prime }}^{A}\psi _{AB}=0  \label{e32}
\end{equation}%
and%
\begin{equation}
\nabla _{A}^{B^{\prime }}\psi ^{AB}-\frac{1}{2}m^{2}A^{BB^{\prime }}+i\beta
_{A}^{B^{\prime }}\psi ^{AB}=0,  \label{e33}
\end{equation}%
together with the complex conjugates of (\ref{e32}) and (\ref{e33}). A
formal simplification to it can be accomplished by utilizing Eqs. (\ref{e24}%
) along with metric prescriptions of the type%
\begin{equation}
\nabla ^{AB^{\prime }}\psi _{AB}=\nabla ^{AB^{\prime }}(\psi _{A}^{C}\gamma
_{CB}),\text{ }\nabla _{AB^{\prime }}\psi ^{AB}=\nabla _{AB^{\prime
}}(\gamma ^{BC}\psi _{C}^{A}).  \label{e34}
\end{equation}%
For the unprimed wave functions, for instance, we thus have the equivalent
statements%
\begin{equation}
\nabla ^{AB^{\prime }}\psi _{A}^{B}+\frac{1}{2}m^{2}A^{BB^{\prime }}=0,\text{
}\nabla _{AB^{\prime }}\psi _{B}^{A}-\frac{1}{2}m^{2}A_{BB^{\prime }}=0.
\label{e35}
\end{equation}%
Evidently, the symmetry borne by the wave functions makes it immaterial to
order their indices.

\section{Wave equations}

At this stage, we shall follow up the procedure which amounts to
implementing the calculational techniques mentioned in Section 1 towards
deriving the wave equations for the fields that occur in the statements (\ref%
{e32})-(\ref{e35}). We will initially work out the procedure for $\psi ^{AB}$
and $\psi _{A}^{B}$. The wave equation for $\psi _{AB}$ will then be
obtained by taking into effect a valence interchange rule that had been
deduced originally [30] in connection with the presentation of the general
description of $\gamma \varepsilon $-curvatures. We may certainly get the
wave equations for any primed fields by taking complex conjugates. Equations
(\ref{e24}) will be used so many times in what follows that we will no
longer refer to them explicitly.

We start by operating with $\nabla _{CB^{\prime }}$ on the configuration of
Eq. (\ref{e33}). Hence, using the operator correlation%
\begin{equation}
\nabla _{CB^{\prime }}\nabla _{A}^{B^{\prime }}=\gamma _{LC}\nabla
_{B^{\prime }}^{L}(\gamma _{MA}\nabla ^{MB^{\prime }})=i\beta _{CB^{\prime
}}\nabla _{A}^{B^{\prime }}+\gamma _{LC}\gamma _{MA}\nabla _{B^{\prime
}}^{L}\nabla ^{MB^{\prime }},  \label{e36}
\end{equation}%
together with the splitting%
\begin{equation}
\nabla _{B^{\prime }}^{L}\nabla ^{MB^{\prime }}=\Delta ^{LM}+\frac{1}{2}%
\gamma ^{LM}\square  \label{e37}
\end{equation}%
and the definition\footnote{%
The object $\square $\ equals the covariant D'Alembertian operator $\nabla
^{\mu }\nabla _{\mu }$ whilst $\Delta ^{AB}$ is linear and enjoys the
Leibniz rule property.}%
\begin{equation}
\Delta ^{AB}=\nabla _{C^{\prime }}^{(A}\nabla ^{B)C^{\prime }},  \label{e38}
\end{equation}%
we get the contribution 
\begin{equation}
\nabla _{CB^{\prime }}\nabla _{A}^{B^{\prime }}\psi ^{AB}=i\beta
_{CB^{\prime }}\nabla _{A}^{B^{\prime }}\psi ^{AB}+(\Delta _{AC}-\frac{1}{2}%
\gamma _{AC}\square )\psi ^{AB}.  \label{e39}
\end{equation}%
The $\Delta $-derivative of (\ref{e39}) reads%
\begin{equation}
\Delta _{AC}\psi ^{AB}=\frac{R}{6}\gamma _{CA}\psi ^{AB}+\Psi
_{AMC}{}^{B}\psi ^{AM}-2i\phi _{AC}\psi ^{AB},  \label{e40}
\end{equation}%
where $\phi _{AB}$ stands for a wave function for Infeld-van der Waerden
photons and $\Psi _{ABCD}$ is a wave function for gravitons in $\mathfrak{M}$%
. For the $\beta $-term of (\ref{e39}), we have%
\begin{equation}
i\beta _{CB^{\prime }}\nabla _{A}^{B^{\prime }}\psi ^{AB}=i(\beta
_{B^{\prime }(A}\nabla _{C)}^{B^{\prime }}-\frac{1}{2}\gamma _{AC}\beta
^{\mu }\nabla _{\mu })\psi ^{AB}.  \label{e41}
\end{equation}%
In addition, recalling the unprimed relation of (\ref{e23}), produces the
following expansion for the differential kernel of the operated mass term
coming from (\ref{e33}):%
\begin{equation}
\nabla _{CB^{\prime }}A^{BB^{\prime }}=\gamma _{MC}(\psi ^{MB}+\frac{1}{2}%
\gamma ^{MB}\nabla _{\mu }A^{\mu }),  \label{e42}
\end{equation}%
which, by virtue of Eq. (\ref{e15}), may be simplified to%
\begin{equation}
\nabla _{CB^{\prime }}A^{BB^{\prime }}=\gamma _{MC}\psi ^{MB}.  \label{e43}
\end{equation}

We have next to allow for the contribution%
\begin{equation}
\nabla _{CB^{\prime }}(i\beta _{A}^{B^{\prime }}\psi ^{AB})=(i\nabla
_{CB^{\prime }}\beta _{A}^{B^{\prime }})\psi ^{AB}+i\beta _{A}^{B^{\prime
}}\nabla _{CB^{\prime }}\psi ^{AB}.  \label{e44}
\end{equation}%
It is evident that the sum of the $\beta $-term of (\ref{e39}) with the
second term lying on the right-hand side of (\ref{e44}), bears skewness in
the indices $A$ and $C$, that is to say,%
\begin{equation}
i\beta _{CB^{\prime }}\nabla _{A}^{B^{\prime }}\psi ^{AB}+i\beta
_{A}^{B^{\prime }}\nabla _{CB^{\prime }}\psi ^{AB}=i\gamma _{CA}\beta _{\mu
}\nabla ^{\mu }\psi ^{AB}.  \label{Add1}
\end{equation}%
For the other individual term of (\ref{e44}), we spell out the auxiliary
configurations%
\begin{equation}
i\nabla _{B^{\prime }[C}\beta _{A]}^{B^{\prime }}=\frac{1}{2}\gamma
_{AC}(\beta ^{\mu }\beta _{\mu }-i\nabla _{\mu }\beta ^{\mu })  \label{e45}
\end{equation}%
and%
\begin{equation}
i\nabla _{B^{\prime }(C}\beta _{A)}^{B^{\prime }}=i(\Delta _{AC}\Phi +2\phi
_{AC}),  \label{e46}
\end{equation}%
where $\Phi $ is given by Eq. (\ref{e25}). The $\Delta $-derivative of (\ref%
{e46}) vanishes identically\footnote{%
Within the $\gamma \varepsilon $-framework, the quantity $\Phi $ is looked
upon as a world scalar subject to a suitable gauge behaviour.} because of
the torsionlessness of $\nabla _{\mu }$. It follows that, fitting pieces
together, yields%
\begin{equation}
(\square +2i\beta ^{\mu }\nabla _{\mu }+\Theta +\frac{R}{3}+m^{2})\psi
^{AB}-2\Psi ^{AB}{}_{LM}\psi ^{LM}=0,  \label{e47}
\end{equation}%
with%
\begin{equation}
\Theta \doteqdot -\beta ^{\mu }\beta _{\mu }+i\nabla _{\mu }\beta ^{\mu }.
\label{e48}
\end{equation}

The entire derivation of the wave equation for the field involved in Eqs. (%
\ref{e35}) does not produce any couplings other than a gravitational one
which looks like that borne by (\ref{e47}). Roughly speaking, the only
reason for this rests upon the result that we can carry out the relevant
derivation without having to call for any correlations like (\ref{e36}) or (%
\ref{e45}), with the valence pattern of $\psi _{A}^{B}$ accordingly ensuring
the absence of any $\phi \psi $-interactions.\footnote{%
The work of Ref. [30] describes in detail on the basis of the theory of spin
densities the situation related to the eventual absence of electromagnetic
contributions from $\Delta $-derivatives.} For the first of Eqs. (\ref{e35}%
), say, we thus reexpress (\ref{e37}) as%
\begin{equation}
\nabla _{CB^{\prime }}\nabla ^{AB^{\prime }}=\gamma _{MC}(\Delta ^{AM}-\frac{%
1}{2}\gamma ^{AM}\square ),  \label{e49}
\end{equation}%
and let the splitting (\ref{e49}) act on $\psi _{A}^{B}$ such that the
relation (\ref{e43}) still holds. Consequently, by taking account of the
derivative%
\begin{equation}
\Delta _{C}^{A}\psi _{A}^{B}=\frac{R}{6}\psi _{C}^{B}+\Psi ^{AB}{}_{CD}\psi
_{A}^{D},  \label{e50}
\end{equation}%
while resetting the kernel for the mass term as%
\begin{equation}
\nabla _{CB^{\prime }}A^{BB^{\prime }}=\psi _{C}^{B},  \label{e51}
\end{equation}%
and making some index substitutions thereafter, we obtain%
\begin{equation}
(\square +\frac{R}{3}+m^{2})\psi _{A}^{B}{}+2\Psi _{AD}{}{}^{BC}\psi
_{C}^{D}{}=0.  \label{e52}
\end{equation}%
It is worth pointing out that the $\Delta $-derivative of (\ref{e50})
possesses the property%
\begin{equation}
\Delta ^{A[C}\psi _{A}^{B]}{}=0.  \label{e53}
\end{equation}

The wave equation for $\psi _{AB}$ can indeed be derived from (\ref{e47}) by
applying to it the simultaneous interchanges%
\begin{equation}
i\beta ^{\mu }\nabla _{\mu }\leftrightarrow -i\beta ^{\mu }\nabla _{\mu },%
\text{ }\Theta \leftrightarrow \overline{\Theta },  \label{e54}
\end{equation}%
which come naturally from the utilization of the differential devices%
\begin{equation}
\square \psi _{AB}=\square (\psi {}^{CD}\gamma _{CA}\gamma _{DB}),\text{ }%
\square (\gamma _{CA}\gamma _{DB})=-\overline{\Upsilon }\gamma _{CA}\gamma
_{DB}  \label{e55}
\end{equation}%
and 
\begin{equation}
2(\nabla _{\mu }\psi {}^{CD})\nabla ^{\mu }(\gamma _{CA}\gamma
_{DB})=4(2\beta ^{\mu }\beta _{\mu }+i\beta ^{\mu }\nabla _{\mu })\psi _{AB},
\label{e56}
\end{equation}%
with the definition%
\begin{equation}
\Upsilon \doteqdot 2(\beta ^{\mu }\beta _{\mu }-\overline{\Theta }).
\label{e57}
\end{equation}%
We thus have%
\begin{equation}
(\square -2i\beta ^{\mu }\nabla _{\mu }+\overline{\Theta }{}+\frac{R}{3}%
+m^{2})\psi _{AB}-2\Psi {}_{AB}{}^{LM}\psi _{LM}=0.  \label{e58}
\end{equation}

\section{Concluding remarks}

We saw that the coupling $2i\phi _{AC}\psi ^{AB}$ occurs through the
expansions (\ref{e40}) and (\ref{e46}) in the derivation that leads to the
wave equation (\ref{e47}), but it nevertheless turns out to be cancelled
when the derivation is actually carried through. If we had instead worked
out the derivation procedure for $\psi _{AB}$, then such a $\phi \psi $%
-coupling would have once again arisen at some intermediate calculational
steps as can clearly be seen from the combined configurations%
\begin{equation*}
\left( 2\Delta ^{AC}-\gamma ^{AC}\square \right) \psi _{AB}+m^{2}\nabla
_{B^{\prime }}^{C}A_{B}^{B^{\prime }}=2i\nabla _{B^{\prime }}^{C}(\beta
^{AB^{\prime }}\psi _{AB}),
\end{equation*}%
\begin{equation*}
2\Delta ^{AC}\psi _{AB}{}=\frac{R}{3}\psi _{B}^{C}{}{}{}{}-2\Psi
_{B}{}{}^{CMN}\psi _{MN}{}+4i\phi ^{AC}{}\psi _{AB}{}
\end{equation*}%
and%
\begin{equation*}
(\nabla _{B^{\prime }}^{(C}\beta ^{A)B^{\prime }})\psi _{AB}=\left( \Delta
^{AC}\Phi +2\phi ^{AC}\right) \psi _{AB},\text{ }\nabla _{B^{\prime
}}^{[C}\beta ^{A]B^{\prime }}=\frac{1}{2}\gamma ^{CA}\nabla _{\mu }\beta
^{\mu }.
\end{equation*}%
Thus, by using the prescription%
\begin{equation*}
\nabla _{CB^{\prime }}A_{B}^{B^{\prime }}=\nabla _{CB^{\prime }}(\gamma
^{B^{\prime }C^{\prime }}A_{BC^{\prime }}),
\end{equation*}%
likewise invoking one of the relationships (\ref{e22}) and implementing (\ref%
{e15}), we could rearrange the kernel of the differentiated mass term of Eq.
(\ref{e32}) as%
\begin{equation*}
\nabla _{CB^{\prime }}A_{B}^{B^{\prime }}=\psi _{BC}+i\beta _{CB^{\prime
}}A_{B}^{B^{\prime }},
\end{equation*}%
which particularly carries the potential coupling%
\begin{equation*}
i\Phi _{CB^{\prime }}A_{B}^{B^{\prime }}=i(\Phi _{B^{\prime
}(B}A_{C)}^{B^{\prime }}-\frac{1}{2}\gamma _{BC}\Phi _{\mu }A^{\mu }).
\end{equation*}%
In fact, the desirable covariance of Eq. (\ref{e58}) under the geometrically
intrinsic $\gamma $-formalism gauge transformation [30]%
\begin{equation*}
\Phi _{\mu }\mapsto \Phi _{\mu }-\partial _{\mu }\theta ,
\end{equation*}%
is brought about when we call for the contribution%
\begin{equation*}
2i\beta ^{AB^{\prime }}\nabla _{CB^{\prime }}\psi _{AB}=(2i\beta ^{\mu
}\nabla _{\mu }+\beta ^{\mu }\beta _{\mu })\psi _{BC}+im^{2}\beta
_{CB^{\prime }}A_{B}^{B^{\prime }},
\end{equation*}%
which accordingly entails the cancellation of all $\Phi $-potential
couplings.

It should be obvious that both of the wave equations (\ref{e47}) and (\ref%
{e58}) could be readily derived from (\ref{e52}) by taking into account the
correlations%
\begin{equation*}
(\square \psi _{A}^{C}{})\gamma _{CB}=(\square -2i\beta ^{\mu }\nabla _{\mu
}+\overline{\Theta })\psi _{AB}{}
\end{equation*}%
and%
\begin{equation*}
\gamma ^{AC}(\square \psi _{C}^{B}{})=(\square +2i\beta ^{\mu }\nabla _{\mu
}+\Theta )\psi ^{AB},
\end{equation*}%
along with the eigenvalue equations%
\begin{equation*}
\square \gamma _{AB}=\Theta \gamma _{AB},\text{ }\square \gamma ^{AB}=%
\overline{\Theta }\gamma ^{AB}.
\end{equation*}

The work we have just presented has provided us with the characteristic
patterns of the $\gamma $-formalism version of the theory of classical Proca
fields. We emphasize that one of the most remarkable properties of the wave
equations deduced previously, is that the only interactions carried by them
involve Proca fields and gravitational wave functions. Hence, Proca fields
propagate in $\mathfrak{M}$ as if Infeld-van der Waerden electromagnetic
curvatures were absent. Therefore, we could say that our work has filled in
the gap associated to the absence of a formal two-component description of
external massive spin-one fields in general relativity. We believe that it
would be of considerable interest to obtain the physically meaningful
couplings involving external spinning fields, which should arise within the
torsional framework of Refs. [35, 36].

ACKNOWLEDGEMENTS

One of us (JGC) should acknowledge the referees for producing many
improvements on the paper. The work carried out here was supported in part
by the Brazilian agency CAPES.

\end{document}